\documentclass[11pt]{article}
\pdfoutput=1
\usepackage{jheppub}
\usepackage{amsmath}
\usepackage{bm}
\usepackage{slashed}
\usepackage{graphicx}

\newlength{\dummysp}
\newcommand{\beq}{\begin{eqnarray}}
\newcommand{\eeq}{\end{eqnarray}}

\newcommand{\gappeq}{\mathrel{\rlap {\raise.5ex\hbox{$>$}}
{\lower.5ex\hbox{$\sim$}}}}
\newcommand{\lappeq}{\mathrel{\rlap{\raise.5ex\hbox{$<$}}
{\lower.5ex\hbox{$\sim$}}}}

\newcommand{\ben}{\begin{enumerate}}
\newcommand{\een}{\end{enumerate}}

\newcommand{\bit}{\begin{itemize}}
\newcommand{\eit}{\end{itemize}}

\def\[{\left [}
\def\]{\right ]}
\def\({\left (}
\def\){\right )}
\def\R{{\mathbb R}}

\def\Z{{\mathbb Z}}

\title{Anomaly matching, (axial) Schwinger models, and high-T super Yang-Mills domain walls}
   
 \author[a]{Mohamed M. Anber,}\author[b]{Erich Poppitz} 
  
\affiliation[a]{Department of Physics, Lewis $\&$ Clark College, Portland, OR 97219, USA}
\affiliation[b]{Department of Physics,   University of Toronto, 
Toronto, ON M5S 1A7, Canada}
\emailAdd{manber@lclark.edu}\emailAdd{poppitz@physics.utoronto.ca}    
\abstract{ 
We study the discrete chiral- and center-symmetry 't  Hooft anomaly matching in the charge-$q$ two-dimensional Schwinger model. We show that the algebra of the discrete symmetry operators involves a central extension, implying the existence of $q$  vacua, and that the chiral and center symmetries are spontaneously broken. 
We then argue that an axial version of the $q$$=$$2$  model appears in the worldvolume theory on domain walls between center-symmetry breaking vacua in the  high-temperature $SU(2)$ ${\cal N}$$=$$1$ super-Yang-Mills theory and that it inherits  the discrete 't Hooft anomalies of the  four-dimensional bulk. The Schwinger model results suggest that the high-temperature domain wall exhibits a  surprisingly rich structure: it supports a non-vanishing fermion condensate and perimeter law for spacelike Wilson loops,  thus mirroring many properties of the  strongly coupled four-dimensional low-temperature theory. We  also discuss  generalizations to theories with multiple adjoint fermions and possible lattice tests.}

\begin{document}

\maketitle

\flushbottom

\section{Introduction}

Quantum field theory (QFT) is a universal paradigm for writing down the fundamental laws of nature.  In many situations, however, QFTs are strongly coupled and learning about their nonperturbative behavior becomes a daunting task. One of the powerful tools that sheds light on the nonperturbative structure of QFT is 't Hooft's anomaly matching  \cite{tHooft:1980xss}. Given a QFT with a continuous or discrete global symmetry $G$,  one may try to introduce a background gauge field of $G$. If the theory doesn't maintain its gauge invariance, we say that it has a 't Hooft anomaly. The anomaly is renormalization group invariant and must be matched between the infrared (IR) and ultraviolet (UV) dynamics. This matching is especially powerful in asymptotically free theories: one computes the anomaly coefficient upon gauging $G$ in the UV, where the theory is amenable to perturbative analysis. Then, this coefficient has to be matched in the IR, which puts constraints on the strongly coupled IR spectrum of the theory, see \cite{Frishman:1980dq,Csaki:1997aw}. If $G=G_1\times G_2$, then it might happen that $G_1$ and $G_2$ have no 't Hooft anomalies, but the product  $G_1\times G_2$ is anomalous. In this case, we say that the theory has a mixed 't Hooft anomaly. 

A global symmetry $G$ is said to be a $0$-form symmetry if it acts on local operators. If $G$ acts on operators of spacetime dimension $q$, then $G$ is a $q$-form symmetry \cite{Gaiotto:2014kfa}. A famous example is  $SU(N)$ Yang-Mills theory, which enjoys a $1$-form $\mathbb Z_N^C$ center symmetry that acts on Wilson line operators. Recently, it has been realized that gauging the $1$-form discrete symmetries can also be obstructed due to the existence of 't Hooft anomalies, which can provide more handles to study the phases of gauge theories \cite{Gaiotto:2014kfa,Gaiotto:2017yup,Sulejmanpasic:2018upi,Tanizaki:2018xto,Shimizu:2017asf,Tanizaki:2017qhf}. In particular, non trivial constraints can be imposed on the vacua of gauge theories (including their number)  that enjoy both $0$- and $1$-form discrete symmetries upon gauging the latter. 

With the help of the recently discovered mixed 't Hooft anomaly,  we scrutinize the domain walls in hot super Yang-Mills theory and its multi-adjoint nonsupersymmetric generalizations, describe their worldvolume effective field theory, and examine the bulk-domain wall anomaly inflow.  We begin the paper with  a warm-up exercise: we study the charge-$q$ two-dimensional (2d) vector Schwinger model and uncover the rich structure of its vacuum. This theory has a $1$-form $\mathbb Z_q^{C}$ (for $q\geq 2$) center symmetry along with a $0$-form $\mathbb Z_{2q}^{d\chi}$ anomaly free discrete chiral symmetry. Performing a global discrete chiral transformation in the background of a $\mathbb Z_q^{C}$ $2$-form  gauge field multiplies the partition function by a non-trivial phase $e^{i\frac{2\pi}{q}}$, indicating the existence of a mixed $\mathbb Z_q^{C}$-$\mathbb Z_{2q}^{d\chi}$ 't Hooft anomaly.  A consequence of this anomaly is that the ground state of the system is far from trivial. We exploit the fact that the Schwinger model is exactly solvable  and explicitly construct its Hilbert space,  generalizing \cite{Manton:1985jm,Iso:1988zi} to $q\ge 2$. We show that, in the quantum theory, the  algebra of operators representing the $\mathbb Z_q^{C}$ and $\mathbb Z_{2q}^{d\chi}$ discrete symmetries is modified by a central extension, signaling the presence of a  mixed anomaly, as in \cite{Gaiotto:2017yup}. Further,
  we find that the charge-$q$ model admits $q$ distinct vacua, which saturate  the mixed 't Hooft anomaly. The theory has a non vanishing fermion bilinear condensate and  screens  arbitrary strength electric charges \cite{Iso:1988zi}, implying that both $\mathbb Z_{2q}^{d\chi}$  and  $\mathbb Z_q^{C}$ are broken in the IR. 

In the second part of the paper, we argue that the axial version of the charge-$2$ Schwinger model appears in the worldvolume theory of the domain wall (DW) in the high temperature 4d $SU(2)$ ${\cal N}=1$ super Yang-Mills (SYM) theory.  The 4d theory has an anomaly-free $0$-form $\mathbb Z_{4 }^{d\chi}$ discrete chiral (or $R$-) symmetry and a $1$-form $\mathbb Z_{2}^C$ center symmetry, with a mixed $\mathbb Z_{4}^{d\chi}$-$[\mathbb  Z_2^{C}]^2$  't Hooft anomaly. Anomaly-inflow arguments imply that this anomaly appears on the DW worldvolume as  a  mixed $\mathbb Z_{4}^{d\chi}$-$ \mathbb  Z_2^{C}$ anomaly  \cite{Gaiotto:2014kfa,Gaiotto:2017yup,Gaiotto:2017tne}. Owing to the 2d nature of Schwinger model, its axial version can be mapped to the  charge-$2$ vector model studied in the first part of the paper. The results there show that  the DW inherits the bulk anomaly, which is saturated by the presence of two degenerate vacua with broken chiral and center symmetries. 
Thus, the $q$$=$$2$ Schwinger model results suggest  that the 2d theory on the high-$T$ DW  mirrors many of the properties of the strongly-coupled 4d theory at low-$T$: \begin{enumerate}
\item  The broken $\mathbb Z_{4}^{d \chi}$ chiral symmetry on the wall implies that the fermion bilinear condensate should be nonzero on the DW in the high-$T$ chirally restored  phase, something that should be in principle measurable on the lattice. 
In some sense, as far as the chiral symmetry is concerned, the high-$T$ dynamics on the DW  resembles the low-$T$ dynamics of the bulk, where  SYM  theory is known to have two vacua with a broken discrete chiral (or $R$-) $\mathbb Z_{4}^{d \chi}$ symmetry. 
\item The broken $\mathbb Z_2^C$ one-form center symmetry on the high-$T$ DW  implies a perimeter law for a fundamental Wilson loop taken to lie in the DW worldvolume. In contrast,  Wilson loops in the $\R^3$ bulk away from the DW exhibit area law (or unbroken $1$-form center symmetry). 
Here, we see again that the DW theory reflects properties of the low-$T$ phase: the different behavior of the Wilson loop in the bulk and on the DW
 mirrors the deconfinement of quarks on the DWs (i.e. perimeter law for the Wilson loop along the DW) between chiral-breaking vacua in the confined low-$T$ phase (i.e. area law in the bulk)  as observed in \cite{Anber:2015kea}, see also \cite{Komargodski:2017smk}.
\end{enumerate}
We find these correspondences between high-$T$ DW physics and low-$T$ bulk and DW physics  quite fascinating.
The matching of various anomalies  and the rich  DW physics uncovered make these properties  worth pointing out and pursuing further.\footnote{The  spirit of the correspondences outlined resembles  those found in the high-$T$ DWs of pure Yang-Mills theory at $\theta = \pi$  \cite{Gaiotto:2017yup}  but the dynamics here appears richer.}
 
This paper is organized as follows. In Section \ref{Discrete 't Hooft anomalies}, we study the charge-$q$ Schwinger model, its discrete symmetries, its 't Hooft anomalies, and the anomaly saturation. In Section \ref{su2dw}, we review the DW solution in the high temperature $SU(2)$ SYM theory and show that the worldvolume of the DW is a  charge-$2$ axial Schwinger model. We also discuss the anomaly inflow and the manifestation of the anomaly on the DW. We conclude, in Section \ref{proposed}, by a discussion of the generalizations to QCD(adj) with a larger number of adjoint fermions and a  proposal to study the high-$T$ domain walls on the lattice.

\section{Discrete 't Hooft anomalies in the charge-$\mathbf q$ Schwinger model }
\label{Discrete 't Hooft anomalies}
Consider the charge-$q$ vector massless  Schwinger model with Lagrangian\footnote{The charge-$q$ Schwinger model was also discussed in \cite{Hansson:1994ep}, but with no reference to anomalies.}
\beq
\label{lagr1}
L = - {1 \over 4 e^2} \;f_{kl} f^{kl} + i \bar\psi_+  (\partial_- + i q A_-) \psi_+  + i \bar\psi_-  (\partial_+ + i q A_+) \psi_-~,
\eeq
where   $k,l=0,1$ are spacetime indices, $\partial_\pm \equiv \partial_t \pm \partial_x$, $A_\pm \equiv A_t \pm A_x$, $t$ and $x$ are the two-dimensional Minkowski space coordinates, $q \ge 2$ is an integer and $e$ is the gauge coupling. The spacetime metric is $g^{kl}=$ diag$(+,-)$, and we further assume that space is compactified on a circle of circumference $L$, with $x \equiv x+ L$. The fields $\psi_+$ ($\psi_-$) are the left (right) moving components of the Dirac fermion and $\bar\psi_\pm$ are the hermitean conjugate fields. Our notation follows from that of  \cite{Iso:1988zi} and, as in that reference, we impose antiperiodic boundary conditions on $\psi_\pm$ around the spatial circle.\footnote{We note that we could also follow \cite{Manton:1985jm} and take the fermions periodic, with no change in the results regarding symmetry realizations and anomalies; also, the utility of Weyl fermion notation  will become clear further below.}

The major difference of our discussion from that in \cite{Manton:1985jm,Iso:1988zi}---where the model (\ref{lagr1}) with $q=1$ was solved exactly in Hamiltonian language for arbitrary values of $L$ (see also the textbook \cite{Shifman:2012zz} which emphasizes the $e L \ll 1$ limit)---is in the assumption that $q>1$ and in the corresponding global issues and discrete anomalies that arise.\footnote{We caution the reader against concluding that the value of $q$ is irrelevant: we are considering a compact $U(1)$ theory with (light) dynamical charges with quantized charge $q>1$. The theory can be probed with nondynamical $q=1$ charges. One can think of the latter as of very (infinitely) massive dynamical charges.}
Understanding the symmetry structure and anomalies of  (\ref{lagr1}) is of interest from multiple points of view:
\begin{enumerate}
\item On its own, the charge-$q$ vectorlike Schwinger model (\ref{lagr1})
 is an interesting example that   provides an exactly solvable setting to study the manifestation of the recently discovered mixed discrete $0$-form/$1$-form 't Hooft anomalies \cite{Gaiotto:2014kfa,Gaiotto:2017yup}. 
\item Two-dimensional models closely related to (\ref{lagr1}) also appear within the framework of four-dimensional gauge theories. We show in Section~\ref{su2dw} that the axial version of the Schwinger model (\ref{lagr1}) with $q = 2$ arises as a worldvolume theory on domain walls (DWs) between center-symmetry breaking states in high-temperature  $SU(2)$ super-Yang-Mills theory, i.e. Yang-Mills theory  with $n_f = 1$ adjoint Weyl fermions. 

Similarly, related multi-flavor axial generalizations of (\ref{lagr1}) appear as worldvolume theories on hot DWs in $SU(2)$ gauge theories with $n_f > 1$ adjoint Weyl fermions. 
\item It turns out that, in all cases mentioned above, the $0$-form/$1$-form 't Hooft anomalies lead to a  rich structure of the DWs that is in principle amenable to lattice studies. As opposed to the study of $\theta=\pi$  pure Yang-Mills theories, where related anomalies arise \cite{Gaiotto:2017yup,Gaiotto:2017tne,Komargodski:2017dmc}, the sign problem does not hinder the lattice studies of these theories (in the continuum limit \cite{Bergner:2018unx}), at least for real values of the fermion mass (of course, here the chiral limit will have to be approached).  A proposal for such studies will be discussed in Section~\ref{proposed}. 
\end{enumerate}

\subsection{Symmetries and mixed 't Hooft anomaly}
\label{mixed}
Thus armed with reasons to study the symmetries and dynamics of (\ref{lagr1}), we proceed to the salient points. We begin with a discussion of the symmetries of the  model (\ref{lagr1}). In addition to the gauged vectorlike symmetry $U(1)_V$, under which $\psi_\pm \rightarrow e^{i q \alpha} \psi_{\pm}$, the model has an anomalous global axial $U(1)_A$ symmetry: 
\beq
\label{axialU1}
U(1)_A: ~ \psi_\pm \rightarrow e^{\pm i \chi} \psi_\pm,   ~{\rm with~ anomaly ~ free~ subgroup } ~ ~{\mathbb{Z}_{2q}^{d \chi}}: ~ \psi_\pm \rightarrow e^{\pm i { \pi \over   q}} \psi_\pm  ~.
\eeq Under a $U(1)_A$ transformation, the fermion measure changes by a factor of $e^{ i 2 q \chi T}$, where $T =  {1 \over 2 \pi}  \int f_{12} d^2 x \in \Z$ is the integer topological charge of the gauge field; recall that we allow probes with $q=1$ and note that we temporarily adopted Euclidean notation. Thus, for $q \ge  2$, a discrete $\mathbb{Z}_{2q}^{d \chi}$ subgroup of the $U(1)_A$ axial transformations, the anomaly free discrete chiral symmetry, survives. 
Under the discrete chiral symmetry $\psi_\pm$ transform   with $\chi = {2 \pi \over 2 q}$, as also indicated on the r.h.s. of (\ref{axialU1}). Notice that for $q=1$ there is only a fermion number symmetry and no nontrivial chiral symmetry. The ${\mathbb{Z}_{2q}^{d \chi}}$ symmetry (\ref{axialU1}) is a $0$-form symmetry as it  acts on the local  degrees of freedom.

A further global symmetry of the $q \ge 2$ theory is the $1$-form $\mathbb Z_q^{C}$ center symmetry. It does not act on any local degrees of freedom, but only on line operators, as its name suggests.\footnote{This is easiest to understand on the lattice, where the global $\mathbb Z_q^{C}$ center symmetry acts by multiplying the unitary  links representing the gauge field component in the $\hat\mu$-direction  by a $\hat\mu$-dependent $\mathbb Z_q$ phase factor, very much as  in (\ref{center}). Thus, the symmetry parameter itself is a $\mathbb Z_q$ valued link, or a $1$-form; see \cite{Greensite:2011zz,Kapustin:2014gua,Gaiotto:2014kfa} for a variety of perspectives.} The $\mathbb Z_q^{C}$ $1$-form center symmetry action on the Wilson loop around the spatial circle, $W \equiv e^{ i \oint  A_x d x}$, is to multiply it by a $\mathbb Z_q$ phase factor 
\beq
\label{center}
\mathbb Z_q^{C}: ~ e^{ i \oint  A_x d x} ~  \rightarrow~ \omega_q \; e^{ i \oint  A_x d x},~~ \omega_q \equiv e^{i\; {2 \pi \over q}}.
\eeq

Both the chiral $0$-form and center $1$-form discrete symmetries, (\ref{axialU1}) and (\ref{center}), are exact symmetries of the quantum theory. However,   they suffer a 't Hooft anomaly:  gauging one of the symmetries explicitly breaks the other  so that they can not be simultaneously gauged. Gauging the $1$-form $\mathbb Z_q^{C}$ center symmetry is most straightforward on the lattice:  one  introduces a $2$-form (plaquette-based) $\mathbb Z_q^{C}$ gauge field to make the $1$-form symmetry (acting on links) local.\footnote{In two spacetime dimensions, there is no $3$-form field strength associated to the $2$-form $\mathbb Z_q^{C}$ gauge field, thus any background is necessarily topological, see e.g.~\cite{Kapustin:2014gua}.} 
In continuum language, introducing a $2$-form $\mathbb Z_q^{C}$ gauge field background is equivalent, see discussion in \cite{Greensite:2011zz}, to turning on nontrivial 't Hooft fluxes, known to carry fractional topological charge  $T = {k \over q}$ ($k \in \Z$)  (see \cite{tHooft:1979rtg,tHooft:1981sps}, dimensionally reduced).

Now, as argued in the paragraph after eq.~(\ref{axialU1}),  under a discrete chiral $ {\mathbb{Z}_{2q}^{d \chi}}$ transformation, the fermion measure changes by a phase factor $e^{i  2 \pi  T}$. This factor is unity for integer $T$, but equals $\omega_q = e^{i  {2 \pi \over q}}$ when a fractional topological charge (a nontrivial $2$-form center gauge background with $k=1$) is introduced. The phase in the chiral transformation of the partition function in the  't Hooft flux background  is the manifestation of the mixed $\mathbb Z_{2q}^{d \chi}$-$\mathbb Z_{q}^C$ 't Hooft anomaly. This phase is renormalization group invariant---it is independent of the volume of the spacetime torus and can also be viewed as the variation of a bulk 3d term \cite{Gaiotto:2017yup,Komargodski:2017dmc}. Ref.~\cite{Gaiotto:2014kfa,Gaiotto:2017yup} argued that this anomaly has to be matched by the infrared (IR) dynamics of the theory and outlined various options for the way the matching can happen.

We  show below that the $\mathbb Z_{2q}^{d \chi}$-$\mathbb Z_{q}^C$ mixed 't Hooft anomaly in the $q \ge 2$ Schwinger model  is reproduced by the IR theory in the ``Goldstone" mode such that both the discrete chiral and center symmetries are spontaneously broken. We also explicitly show that the mixed anomaly in the $q \ge 2$ Schwinger model appears as a ``central extension" of the algebra of the operators generating the discrete chiral $\mathbb Z_{2q}^{d \chi}$ and center $ \mathbb Z_{q}^C$
transformations, see Eq.~(\ref{algebra}) in the next Section.\footnote{This is similar to the  appearance of the CP/center anomaly in the quantum mechanical and field theory models of \cite{Gaiotto:2017yup,Komargodski:2017dmc,Kikuchi:2017pcp}.  Enhancement of the discrete symmetry group in Yang-Mills theory at $\theta=\pi$ due to discrete 't Hooft anomaly considerations was also discussed in \cite{Aitken:2018kky}.}

\subsection{The realization of the symmetries and their algebra}

In this Section, we study the realization of the discrete symmetries and their 't Hooft anomaly in the charge-$q$ Schwinger model (\ref{lagr1}), by   borrowing the results of \cite{Iso:1988zi,Manton:1985jm}. As our  focus is on the symmetry realization, we shall be mostly concerned with the properties of the ground state. 
Briefly, the strategy behind the first steps of the Hamiltonian solution of (\ref{lagr1}) in $A_t = 0$ gauge is to explicitly solve  the Weyl equation  in the $A_x$ background  (this is possible in one space dimension) and use its eigenfunctions and eigenvalues   to construct  Dirac sea states. To find the physical ground state, one then imposes  Gauss' law, i.e., invariance under infinitesimal gauge transformations. Finally, one  demands that the vacuum states be eigenstates of the large gauge transformations $G: A_x \rightarrow e^{i g(x)}(A_x + i \partial_x) e^{- i g(x)}$, where $e^{ig(x)} \equiv e^{i {2 \pi x \over L}}$ is the unit winding number large gauge transformation. 
 
To introduce  some of the notation of \cite{Iso:1988zi}, the holonomy of the gauge field around the spatial circle is $
\oint A_x dx \equiv c L, 
$
with $cL$  shifted by $2\pi$ under large gauge transformations $G$. The action of the center symmetry (\ref{center}) on the holonomy $cL$   is
 \beq
\label{center2}
  \mathbb Z_q^{C}: ~ cL \rightarrow cL + {2 \pi \over q}~.
\eeq
The Dirac sea states obeying Gauss' law can be found as was briefly outlined above. The end result is that the states are labeled  by an integer $n$ and we shall simply denote them  by $| n \rangle$, not displaying their dependence on $cL$; the  explicit form is in \cite{Iso:1988zi}. The Dirac sea state $|n\rangle$ is the one where the states of all left moving particles of (gauge non-invariant) momenta $\le {2 \pi (n-1)\over L}$ are occupied and the rest are empty, and, simultaneously,   all states of the right moving particles of momenta $\ge {2 \pi n \over L}$ are occupied. This   left vs. right moving ``Fermi level" matching ensures validity of the Gauss' law  \cite{Iso:1988zi,Manton:1985jm}. 

{\flushleft{W}}e now list the properties of the  Dirac sea  states $|n \rangle$  that matter  to us.  See  \cite{Iso:1988zi} for precise definitions and derivations. We  notice that $q>1$ is easily incorporated and is seen to lead  to important new points, see items \ref{3}, \ref{4}, and \ref{5} below:
\begin{enumerate}
\item The different $|n\rangle$ states are orthogonal; their norm can be defined as unity, $\langle n | m \rangle = \delta_{mn}$.
\item Their $U(1)_V$ charge vanishes, but the chiral (or axial $U(1)_A$, recall (\ref{axialU1})) charge $Q_5$, is nonzero and depends on the holonomy of the gauge field 
\beq
\label{vacQ5}
Q_5 |n \rangle = |n \rangle \left(2 n - {q c L \over \pi}\right)~.
\eeq
The gauge field-dependence of the axial charge $Q_5$ is a reflection of the chiral anomaly. One can define a gauge-field independent $\tilde Q_5$ with integer eigenvalues
\beq
\label{vactildeQ5}
\tilde{Q}_5 \equiv Q_5 + {q c L \over \pi},
\eeq
but this operator shifts under large gauge transformations 
\beq
\label{vactildeQ51}
G: \tilde{Q}_5 \rightarrow \tilde{Q}_5 + 2 q~.
\eeq
\item  \label{3}
It is clear, however, that the operator 
\beq
\label{vactildeQ52}
X_{2q} \equiv e^{i {2 \pi \over 2 q} \tilde Q_5}
\eeq
is invariant under large gauge transformations. It generates the $\mathbb Z_{2q}^{d \chi}$ anomaly free subgroup of the chiral transformations (\ref{axialU1}) and acts on the $|n \rangle$ states as
\beq
\label{chiralaction}
X_{2q} |n \rangle = |n \rangle \; \omega_q^n~~ ~~~~( \omega_ q \equiv e^{i {2 \pi   \over q}} ).
\eeq
\item
The  Dirac sea states $| n \rangle$ are eigenstates of the fermion Hamiltonian $H^F$ in the $A_x$ background  and their energies are
\beq
\label{energy}
E_n^F ={2 \pi \over L} \left[{Q_5^2  \over 4} - {1 \over 12} \right] = 
 {2 \pi \over L} \left[{1 \over 4}\left( 2 n - {q c L \over   \pi}\right)^2 - {1 \over 12} \right]~.
\eeq
(Here and elsewhere we take the liberty to denote operators and eigenvalues with the same letter, hoping that this does not cause undue confusion.) 

As eigenstates of the total Hamiltonian, however, the $|n \rangle$ states, supplemented by a holonomy wave function \cite{Iso:1988zi,Manton:1985jm},  are degenerate; one way to  see this is by noting that the holonomy fluctuations $cL$ obtain the same ``mass" from the fermion vacuum energy (\ref{energy}) in all $|n\rangle$ Dirac sea states.\footnote{Diagonalizing the full Hamiltonian, including fermion excitations above the Dirac sea, in the language used here involves a Bogolyubov transformation  \cite{Iso:1988zi}, but the details will not be relevant for us.}
\item \label{4}
Under large gauge transformations  $G$, shifting $cL$ by $2 \pi$,  the $|n \rangle$ states are not invariant but transform into each other as
 \beq
 \label{vacG}
 G |n \rangle = | n+ q \rangle~.
 \eeq
 \item \label{5}
 The center symmetry $\mathbb Z_q^C$,  a $2 \pi \over q$ shift of $cL$ (\ref{center2}),  acts on the $|n \rangle$ states as
 \beq
 \label{vacC}
Y_q |n \rangle = |n+1 \rangle~,
 \eeq
 where we introduced the  $Y_q$  operator, representing the center-symmetry action on the gauge field holonomy.
 \end{enumerate}
 We are now ready, as in \cite{Iso:1988zi,Manton:1985jm}, to construct states that are eigenstates of the large gauge transformations $G$. Since the $|n\rangle$ states transform as (\ref{vacG}),  in the $q>1$ theory we can define $q$  different linear combinations of the $|n\rangle$ states that are eigenstates of $G$. For convenience, we introduce a $\theta$ parameter (it is unobservable in the massless theory \cite{Manton:1985jm}) and define the linear combinations $| \theta, k \rangle$ of the Dirac sea states as  
 \beq
 \label{thetak}
 | \theta, k \rangle \equiv \sum\limits_{n \in \Z} e^{i (k + q n) \theta} |k+ qn\rangle, ~ k = 0, 1,\ldots,q-1~.
 \eeq
 As follows from (\ref{vacG}), all $| \theta, k\rangle$ states are eigenstates of $G$ with eigenvalue $e^{- i q \theta}$.   We note also that $ \langle \theta', k'| \theta,k\rangle = \delta_{k, k'({\rm mod} \; q)} \;
\delta(\theta - \theta' ({\rm mod} {2 \pi \over q}))$, with $\delta(\theta - \theta' ({\rm mod} {2 \pi \over q}))=\sum\limits_{m\in \Z} e^{i q m (\theta - \theta')}$.
 
 For further use (cluster decomposition, see below), we also  define the   $\Z_q$ Fourier transform of the basis (\ref{thetak}). We denote the states of this basis by $|P, \theta \rangle$ (to not confuse them with the $|\theta, k\rangle$ states):
 \begin{eqnarray}
 \label{thetap}
 | P, \theta\rangle  &\equiv& {1 \over \sqrt{q}} \sum\limits_{k=0}^{q-1} \omega_q^{kP} |\theta, k \rangle, ~ P = 0,\ldots, q-1,  \nonumber \\
  \langle P', \theta' | P, \theta \rangle &=& \delta_{P, P'({\rm mod} \; q)} \;\delta(\theta - \theta' ({\rm mod} {2 \pi \over q})) .
 \end{eqnarray}
Clearly, the   $|P, \theta\rangle$ states are also eigenstates of $G$ with the same eigenvalue $e^{- i q \theta}$. Further,  (\ref{thetap}), (\ref{thetak})  and (\ref{chiralaction}) imply that under the discrete chiral symmetry $\mathbb Z_{2 q}^{d \chi}$ the $| P \rangle$ states transform cyclically into each other
\beq
\label{chip}
X_{2q}\; | P, \theta \rangle = | P + 1 ({\rm mod} \;q), \theta\rangle~,
\eeq
while (\ref{vacC})  implies that they are eigenstates of the $\mathbb Z_{q}^C$ center symmetry 
\beq
\label{center1}
Y_q \; | P, \theta \rangle = |P, \theta \rangle \; \omega_q^{- P} e^{-i \theta}~.
\eeq
Further, following the discussion after (\ref{energy}), the $|P, \theta\rangle$ states are  degenerate. 
The action of $X_{2q}$ and $Y_q$  found above, (\ref{chip}), (\ref{center1}), implies that, when acting on the $| P, \theta \rangle$ states,\footnote{A slightly more careful study of the definitions of the operators from \cite{Iso:1988zi}  shows that the algebra (\ref{algebra}) holds in the entire Hilbert space.} they do not commute but obey the algebra
\beq
\label{algebra}
X_{2q} \; Y_q = \omega_q\; Y_q \; X_{2q} ~ ~ ~ (\omega_q = e^{i {2 \pi \over q}}).  
\eeq
This algebra is familiar from the 't Hooft commutation relation between Wilson and 't Hooft loop operators in $SU(q)$ gauge theories  \cite{tHooft:1977nqb} (the $q$-dimensional representation  on the  $|P,\theta\rangle$ states, (\ref{chip}), (\ref{center1}), was also found there). Here, however, one of the operators $Y_q$, being a center-symmetry generator, is indeed   a (lower dimensional version of a)  't Hooft loop operator, but the other,  $X_{2q}$, is not a Wilson loop but a generator of discrete chiral transformations. 

The 't Hooft algebra (\ref{algebra}) implies that even though the  symmetries generated by $X_{2q}$ and $Y_q$ commute classically, the discrete chiral and center   symmetries $\mathbb Z_{2q}^{d \chi}$ and   $\mathbb Z_{q}^C$  do not  commute in the quantum theory   but instead obey (\ref{algebra}). Their noncommutativity in the quantum theory signals the presence of the  mixed 't Hooft anomaly.\footnote{Following \cite{Gaiotto:2017yup}, we call the appearance of $\omega_q$ in (\ref{algebra}) a ``central extension" of the algebra of symmetry operators, as  the new element $\omega_q$ commutes with $X_{2q}$ and $Y_q$. }

Finally, let us argue that (\ref{algebra})  implies that both symmetries are spontaneously broken. The   $|P, \theta \rangle$ ground states  obey the cluster decomposition principle, as opposed to the $|\theta, k \rangle$ ground states. This is because the  latter are  mixed by local operators, the gauge invariant fermion bilinear  $\phi(x) \equiv \bar\psi_+ (x) \psi_-(x)$. The fermion bilinear has charge $-2$ under the $\mathbb Z_{2q}^{d \chi}$ discrete chiral symmetry (\ref{axialU1}) and nonzero matrix elements between the $|n\rangle$ states:
\beq
\langle n' | \phi(x) |n \rangle  =   \delta_{ n',  n +1}\; C' \; e^{- i {2 \pi x \over L}}\;, ~{\rm where} ~ \phi(x) \equiv \bar\psi_+ (x) \psi_-(x). 
\label{bilinear}
\eeq
The constant $C'$ was computed in   \cite{Iso:1988zi}  in the Hamiltonian formalism for any $L$ and was shown to not vanish,  including as $L \rightarrow \infty$,  where $C' \sim e$. It is also clear that (\ref{bilinear}) is consistent with the nature of the $|n \rangle$ states explained earlier. Using the matrix elements (\ref{bilinear}) it is straightforward to show that $\phi(x)$ has nonzero matrix elements between different $|\theta, k\rangle$ states, $\langle \theta, k+1|\phi|\theta, k\rangle \ne 0$, but  is diagonal in the $|P, \theta \rangle$ basis 
\beq
\label{bilinearP}
\langle P', \theta| \phi(x) | P, \theta \rangle = e^{- i \theta}\;\omega_q^{-P} \; \delta_{P,P'} C',
\eeq
where we took the infinite-$L$ limit on the r.h.s.
Furthermore, one can use Eq.~(\ref{bilinearP}) to show that  correlation functions  factorize \beq\label{cluster1}
\langle P', \theta| \phi^\dagger(x) \phi(0) | P, \theta \rangle\big\vert_{x \rightarrow \infty} \rightarrow  const.\;  \delta_{P,P'} \langle P, \theta| \phi^\dagger(x) |P, \theta \rangle \langle P, \theta | \phi(0) | P, \theta \rangle~,
\eeq
and that connected correlators in the $|P,\theta\rangle$ vacua  vanish as $x \rightarrow \infty$, as required by cluster decomposition. 

To conclude, we have shown that the charge-$q$ Schwinger model has $q$ ground states, $|P, \theta\rangle$, $P=0,1,...q-1$, cyclically permuted by the discrete chiral $\mathbb Z_{2q}^{d\chi}$ symmetry, which is spontaneously broken to $\mathbb Z_2$, see (\ref{bilinearP}).

To see that the $1$-form $\mathbb Z_{q}^C$ symmetry is also broken in the large-$L$ limit, one can study the expectation value of a $q=1$ Wilson loop by  taking two static $q=1$ charges some distance apart and studying the potential between them.  The  calculation of \cite{Iso:1988zi} showed that arbitrary charges are screened in the massless Schwinger model, due to vacuum polarization effects, with  the screening length of order $1/e$ (in the $L\rightarrow \infty$ limit). Hence the Wilson loop obeys perimeter law, signaling the breaking of the $1$-form symmetry \cite{Gaiotto:2014kfa}.

\section{The high-$\mathbf T$ domain wall in  $\mathbf{SU(2)}$ 
super-Yang-Mills: the axial Schwinger model and symmetry realizations}\label{su2dw}

In this Section, we show that the Schwinger model studied above and its generalizations  appear  as the worldvolume theory on high-temperature DWs\footnote{These are  really Euclidean objects, see \cite{Bhattacharya:1990hk,Bhattacharya:1992qb,Smilga:1993vb,KorthalsAltes:1999xb}.} in  4d $SU(2)$ gauge theories with $n_f \le 5$ massless adjoint Weyl fermions, QCD(adj).

 QCD(adj) with $SU(2)$ gauge group  has a discrete anomaly free $\mathbb Z_{4 n_f}^{d\chi}$ $0$-form chiral symmetry and a $\mathbb Z_{2}^C$ $1$-form center symmetry, as well as a continuous $SU(n_f)$ flavor symmetry. The discrete symmetries have a $\mathbb Z_{4 n_f}^{d\chi}$-$\left[\mathbb Z_{2}^C\right]^2$ mixed 't Hooft anomaly 
\cite{Gaiotto:2014kfa,Komargodski:2017smk,Shimizu:2017asf}, much like the one in the Schwinger model of the previous Section. All  't Hooft anomalies have to be matched by the IR dynamics. At zero temperature, the discrete anomaly matching, combined with  continuous and mixed discrete-continuous anomalies, can be used to suggest the possible existence of interesting new  phases \cite{Anber:2018iof}, see the remarks in \cite{Cordova:2018acb}. Furthermore, anomaly inflow arguments for the discrete 't Hooft anomalies imply that there is nontrivial physics on the DWs separating vacua with broken discrete symmetries \cite{Gaiotto:2017yup,Gaiotto:2017tne}. For example, at zero temperature, the discrete chiral symmetry is broken, at least for small $n_f$,  and  DWs connecting different vacua are found to exhibit rich  worldvolume physics  (for example quark deconfinement) \cite{Anber:2015kea} dictated by the anomaly \cite{Gaiotto:2014kfa,Komargodski:2017smk}. 

Here, we focus on the high-temperature regime of $SU(2)$ QCD(adj), in the deconfined phase, where the $\mathbb Z_2^C$ center symmetry  associated with the Euclidean time direction is broken.\footnote{In the terminology of \cite{Gaiotto:2014kfa} the symmetry broken at high-$T$ is called a $0$-form center symmetry, from the point of view of the dimensionally reduced 3d theory. In addition, there is a $1$-form  center symmetry in the dimensionally reduced 3d theory, unbroken at high-$T$ (a fundamental Wilson loop in the spatial $\R^3$ exhibits area law in the deconfined phase).} Here, anomaly inflow arguments also suggest that the DW between the center broken vacua inherits the bulk symmetries and discrete 't Hooft anomalies \cite{Gaiotto:2017yup,Gaiotto:2017tne}. Our goal is to study  this in some detail and see how the 't Hooft anomalies are saturated on the DW worldvolume. We uncover  a rich structure of the DWs  and  find an explicit connection to the Schwinger model of the previous Section  (for $n_f=1$) or its generalizations (for $n_f>1$). 
We begin with the Euclidean action of
 $SU(2)$ Yang-Mills theory endowed with $n_f$ adjoint Weyl fermions at finite temperature $T$:
\begin{eqnarray}
S= \int_{\mathbb R^3\times \mathbb S_\beta^1}\frac{1}{2g^2}\;\mbox{tr}\left( F_{\mu\nu}F_{\mu\nu}\right)+2i\mbox{tr}\left(\bar \lambda \bar \sigma^\mu D_\mu\lambda\right)\,,
\label{SU2 Lagrangian}
\end{eqnarray}
where $\mu,\nu=1,2,3,4$ and the trace is normalized as $\mbox{tr}\left(t^at^b\right)=\frac{\delta_{a,b}}{2}$ such that $t^a=\frac{\tau^a}{2}$ and $\tau^a$  are the color-space Pauli matrices. $\mathbb S_\beta^1$ is the thermal circle, which is taken along the $x_4$-direction and has circumference $\beta=1/T$. The covariant derivative is given by $D_\mu\lambda=\partial_\mu \lambda-i[A_\mu,\lambda]$ and $\bar \sigma=\left(-i,\bm\sigma\right)$, where $\bm \sigma$ are the spacetime Pauli matrices.  In addition, the fermion field $\lambda$ carries an implicit flavor index; we set $n_f=1$ for the rest of this Section.\footnote{\label{higherflavor}The analysis of the zero modes in this Section goes verbatim for any $n_f$, simply increasing the number of zero modes. The analysis of the DW world-volume theory, however, differs for $n_f > 1$, see Sec. \ref{proposed}.}

At temperatures much smaller than the strong coupling scale $\Lambda_{QCD}$, the theory preserves its $\mathbb Z_2^C$ symmetry, the trace of the Polyakov loop vanishes $\mbox{Tr}_F\exp\left[i\oint_{\mathbb S^1_\beta} A_4\right]=0$, static charges are confined, and a Wilson loop wrapped in the time direction obeys the area law.  At temperatures larger than  $\Lambda_{QCD}$, many aspects of the theory become amenable to semiclassical treatment owing to asymptotic freedom. In this regime, we can dimensionally reduce the action (\ref{SU2 Lagrangian}) to $3$d after integrating out a tower of heavy Matsubara excitations of the gauge and fermion fields along $\mathbb S_1^\beta$. To one-loop order, the resulting bosonic part of the action reads 
\begin{eqnarray}
S_{3D}^{{\rm boson}}=\frac{\beta}{g^2}\int_{\mathbb R^3}\left( \frac{1}{2}\mbox{tr}\left( F_{ij}F_{ij}\right)+\mbox{tr}\left(D_i A_4\right)^2+g^2V(A_4)+{\cal O}\left(g^4\right) \right) \,,
\label{bosonic action}
\end{eqnarray}
where $i,j=1,2,3$, $g$  is the gauge coupling at the scale $T$, and $V(A_4)$ is the one-loop effective potential for the Matsubara zero mode of the $x^4$-component of the gauge field. The potential, written below in terms of the Cartan subalgebra component $A_4^3$ and for $n_f =1$, is given (see e.g.~\cite{Anber:2017pak}), up to a constant, by
\begin{eqnarray} \label{one loop} V(A_4)=-\frac{1}{12\pi\beta^4}\left[-6\pi \left(\beta A_4^3\right)^2+4 \left(\beta A_4^3\right)^3\right]\,,\quad {\rm for } ~ \beta A_4^3 \in \left[0,\pi\right]\,,
\end{eqnarray}
where the extension to the interval $\left[\pi, 2\pi\right]$ is given by replacing $\beta A_4^3\rightarrow 2\pi-\beta A_4^3$ in (\ref{one loop}). 
The two minima of the potential are at $\beta A_4^3 = 0, 2\pi$, so that at $T \gg \Lambda_{QCD}$ the $\mathbb Z^C_2$  center symmetry along the $x^4$ direction is broken and the theory admits DWs \cite{Bhattacharya:1990hk,Bhattacharya:1992qb}. The two center-symmetry breaking vacua are characterized by nonvanishing expectation values of the trace of the Polyakov loop, ${1 \over 2} \langle\mbox{Tr}_F\exp\left[i\oint_{\mathbb S^1_\beta} A_4\right]\rangle= \pm 1$.

The DW is a solution of the equations of motion of (\ref{bosonic action}). A DW   perpendicular to $x^3$ is  parameterized as
$
A_\mu^{DW}(x_3)=\delta_{\mu 4}T\Phi(x_3)\frac{\tau^3}{2}\,,
$
where $\frac{\tau^3}{2}$ is $SU(2)$ Cartan generator and the profile function  $\Phi(x_3)$ interpolates between $0$ and $2\pi$, the two $0$-form center-symmetry breaking vacua, as $x_3\rightarrow \mp \infty$. The inverse width of the DW is of order $g   T$  \cite{Bhattacharya:1990hk,Bhattacharya:1992qb}. At the DW core, we find $\Phi(x_3=0)=\pi$ and the trace of the Polyakov  loop vanishes, $\mbox{Tr}_F\exp\left[i\oint_{\mathbb S^1_{\beta}} A_4(x_3=0)\right]=0$, restoring the $0$-form center symmetry on the DW. Furthermore, the center-symmetric expectation value  $\Phi(x_3=0)=\pi$    spontaneously breaks the $SU(2)$ gauge symmetry to $U(1)$ and the off-diagonal $W$-bosons have mass  $\sim T$. 
 Thus, the DW worldvolume supports massless abelian fields. In the $\R^3$ bulk, on the other hand, the gauge sector (\ref{bosonic action}) has a nonperturbative gap, of order $g^2 T$, while on the DW the  $W$-bosons have a larger gap of order $T$, due to the adjoint Higgsing. Thus, in the presence of a DW, we expect that at sufficiently low energy scales (presumably below the bulk gap)  the high-$T$ 3d theory dynamically compactifies to an abelian theory on the   2d worldvolume, in a manner resembling   \cite{Dvali:1996bg}. Having a 2d abelian gauge field on the worldvolume is not very interesting in pure YM theory, except at $\theta=\pi$, where it was shown to have interesting consequences  \cite{Gaiotto:2017yup}. Here, we show that in theories with adjoints the dynamics is even richer. 
 
To this end,  we  study the fermions in the DW background and show that the DW supports two fermionic zero-modes (for $n_f=1$). We expand the gauge fluctuations and fermion fields in Fourier modes, taking into account that the gauge field (fermions) satisfy periodic (anti-periodic) boundary conditions along $\mathbb S^1_\beta$:
\begin{eqnarray}
\nonumber
A_\mu&=&A_\mu^{DW}(x_3)+\sum_{p\in \mathbb Z}\left( a_{\mu,\,p} \frac{\tau_3}{2}+W^+_{\mu,\,p}\tau^++W^-_{\mu,\,p}\tau^-\right)e^{i2\pi p\frac{x_4}{\beta}}\,,\\
\lambda&=&\sum_{p\in \mathbb Z}\left(\lambda^0_{p}\frac{\tau^3}{2}+\lambda^+_{p}\tau^++\lambda^-_{p}\tau^-\right)e^{i2\pi p'\frac{x_4}{\beta}}\,,
\label{Fourier expansion}
\end{eqnarray}
where $p'=p+\frac{1}{2}$ and $\tau^\pm = (\tau^1 \pm i \tau^2)/2$. The photon $a_{\mu\,,p=0}$ is the only massless mode on the DW. All other gauge modes---the W-bosons and their Kaluza-Klein excitations $W^{\pm}_{\mu,\,p}$ as well as the photons $a_{\mu,\,p\neq 0}$---are massive and we neglect them in our treatment. The DW-worldvolume scalar $a_{3,p=0}$ is also expected to be massive as there is no symmetry protecting it, is uncharged under the 2d $U(1)$, and we ignore it in what follows. Substituting (\ref{Fourier expansion}) into the covariant derivative and varying the Lagrangian (\ref{SU2 Lagrangian}) with respect to $\bar \lambda$ we obtain the equation of motion
\begin{eqnarray}
\nonumber
\left(2\pi p'T+\sigma^i\partial_i\right)\lambda_p^0\frac{\tau_3}{2}+\left[\left(2\pi p'-\Phi(x_3)\right)T+\sigma^i\left(\partial_i-ia_{i}\right)\right]\lambda_p^+\tau^+&&\\
 + \left[\left(2\pi p'+\Phi(x_3)\right)T+\sigma^i\left(\partial_i+ia_{i}\right)\right]\lambda_p^-\tau^- &&=0\,,
\label{final EOM}
\end{eqnarray}
where $i,j=1,2,3$ and we used the notation $a_i\equiv a_{i,\,p=0}$ for the Matsubara zero mode of the Cartan gauge field (the $U(1)$ photon). One can immediately see that the lightest $\lambda_p^0$ has a  mass $\sim T$. A careful examination of the (charged under the unbroken $U(1)$) components $\lambda_{p}^{\pm}$, however, reveals that there are two zero modes on the DW. Setting $a_i=0$ and bearing in mind that $\lambda_{p}^{\pm}$ is a two-component spinor, i.e., $\lambda_{p}^{\pm}=\left[\begin{array}{cc}\lambda_{p,\,1}^{\pm}\\ \lambda_{p,\,2}^{\pm} \end{array}\right]$, the $x^3$-dependent solution of the equation of motion of $\lambda_{p}^{\pm}$ reads
\begin{eqnarray}
\nonumber
\lambda^{\pm}_{p,\,1}(x_3)&=&\exp\left[\left(-2\pi p'x^3\pm \int^{x_3}dz\Phi(z)\right)T \right]\lambda^{\pm}_{p,\,1}(0)\,,\\
\lambda^{\pm}_{p,\,2}(x_3)&=&\exp\left[\left(2\pi p'x^3\mp \int^{x_3}dz\Phi(z)\right)T \right]\lambda^{\pm}_{p,\,2}(0)\,.
\end{eqnarray}
It is easy to check that only  two of these solutions, $\lambda^{+}_{p=0,\,2}(x_3)$ and $\lambda^{-}_{p=-1,\,1}(x_3)$, are normalizable. It is crucial for our purposes to note that these two zero modes have opposite charges under the $U(1)$ field $a_i$ and also have opposite 2d chirality, as can be seen from (\ref{final EOM}). 

In what follows, when writing the DW-volume  theory of the zero modes, we  drop the Matsubara and 4d spinor indices, and denote the two dimensional fields corresponding to the above zero modes by $\lambda_+$ and $\lambda_-$, respectively. Also, to emphasize the fact that $\lambda$ are adjoint fermions, and therefore, carry twice the fundamental charge, we make the change of variables $A_{1,2}=\frac{a_{1,2}}{2}$. 
Then, the effective 2d Lagrangian on the DW worldvolume is given by
\begin{eqnarray}
\nonumber
{\cal L}_{DW}^{\rm axial}=\frac{1}{4e^2}F_{kl}F_{kl}&+&i\bar \lambda_+ \left[\partial_1-i\partial_2-  i  2 (a_1-ia_2) \right] \lambda_+ \\
&+&i\bar \lambda_- \left[\partial_1+i\partial_2+ i 2 (a_1+ia_2) \right] \lambda_- \,.
\label{axial Schwinger model}
\end{eqnarray}
where $F_{kl}=\partial_k a_l-\partial_l a_k$, $k,l=1,2$, and $e^2$ is the two dimensional gauge coupling.\footnote{The localization of the abelian  fields on the DW is due to nonperturbative effects  in the bulk that generate a mass gap for the gauge fluctuations (in the  absence of a bulk gap, the abelian gauge field in the DW background would propagate in the $\R^3$ bulk). Thus, we can only estimate the value of the 2d coupling $e^2$: we take $e^2 = g^2 T/\delta$, where $\delta \sim 1/g^2 T$ is the bulk confining scale,  much larger than the DW width, leading  to $e^2\sim g^4 T^2$.  This estimate may raise the issue of scale separation between DW and bulk dynamics: from the above estimate, nonperturbative effects in the 2d Schwinger model occur at scales $e \sim g^2 T$ which is parametrically the same as the nonperturbative bulk gap. These estimates equally apply to the $\theta = \pi$ YM case  of \cite{Gaiotto:2017yup,Gaiotto:2017tne}. 
In what follows, we assume that the  results from Sec.~\ref{Discrete 't Hooft anomalies} apply to the DW theory and offer the heuristic justification that  the only light charged states near the DW are the $\lambda_{\pm}$ zero modes, charged  $W^\pm$-bosons  and fermions have mass of order $T$ on the DW, while  the bulk confined states are uncharged. }
The Lagrangian (\ref{axial Schwinger model}) describes the Euclidean axial Schwinger model of charge $2$ and, from a 4d perspective, the high-$T$ DW worldvolume theory in $SU(2)$ super-Yang-Mills (SYM) theory (QCD(adj) with $n_f=1$). 

 It is interesting to note that the DW worldvolume theory (\ref{axial Schwinger model}) inherits the symmetries and anomalies of the bulk SYM  theory. The $U(1)_A$, under which $\lambda_\pm$ transform with opposite charges, is  gauged  in the axial charge-$2$ model.\footnote{We remind the reader that gauging $U(1)_A$ is possible in 2d, due to the vector-axial  duality ($\epsilon^{\mu\nu}\gamma_\nu = \gamma^\mu\gamma^5$).} The $U(1)_V$, under which $\lambda_\pm$ have the same charge is anomalous, instead.  
There is a $\mathbb Z_{4}^{d \chi}$ discrete ``chiral" (from the bulk point of view) symmetry remaining anomaly free. In addition there is a $\mathbb Z_2^C$ center symmetry due to the fact that the  adjoint fermions carry twice the fundamental charge (this worldvolume $\mathbb Z_2^C$ symmetry originates from the $1$-form center symmetry  in the $\R^3$ bulk and should not be confused with the zero-form center symmetry along $x^4$). There is also a $\mathbb Z_4^{d\chi}$-$\mathbb Z_2^C$ mixed 't Hooft anomaly on the DW worldvolume, as predicted by anomaly inflow \cite{Gaiotto:2017yup,Gaiotto:2017tne}, and as follows directly by repeating the arguments of Section \ref{mixed} for the axial model (\ref{axial Schwinger model}).

To more explicitly see the relation of (\ref{axial Schwinger model}) to the Euclidean version of the vector Schwinger model (\ref{lagr1}), the Lagrangian (\ref{axial Schwinger model})  
 can be brought to the vector form of Schwinger model via the change of variables 
 $\bar\psi_+=\lambda_+$,  $\psi_+=\bar\lambda_+$  in the $+$ sector and a relabeling $\psi_- = \lambda_-$, $\bar\psi_- = \bar\lambda_-$ in the $-$ sector. Performing this in (\ref{axial Schwinger model}), after integration by parts in the $+$ sector we find
\begin{eqnarray}
\label{vectorDW}
\nonumber
{\cal L}_{DW}^{{\rm vector}}=\frac{1}{4e^2}F_{kl}F_{kl}&+& i\bar\psi_+ \left[\partial_1-i\partial_2+ i 2 (a_1-ia_2) \right] \psi_+     \\
&+& i\bar\psi_- \left[\partial_1+i\partial_2+  i  2 (a_1+ia_2) \right] \psi_- 
\end{eqnarray}
 which is the Euclidean version of (\ref{lagr1}) with $q=2$.
Now the anomaly free $\mathbb Z_{4}^{d \chi}$ is a subgroup of the $U(1)_A$ acting on $\psi_+$ and $\psi_-$ with opposite charges, as in (\ref{lagr1}).  We see that at the level of the Lagrangian (\ref{vectorDW}) (where the roles of $U(1)_V$ and $U(1)_A$ are interchanged w.r.t. (\ref{axial Schwinger model})) the symmetries are realized exactly as in the model (\ref{lagr1}).  

For completeness, we should note that the worldvolume theory,\footnote{We stress again that the present discussion applies only to the $n_f=1$ SYM case, see Footnote \ref{higherflavor}. We shall come back to the $n_f>1$ case in the future.} in addition to the terms in (\ref{vectorDW}), allows for a single classically (and quantum mechanically, as  the results of \cite{Sachs:1995dm} imply) marginal coupling of the form 
 \beq
 \label{4fermi}
 {\cal L}_{DW}^{4-{\rm fermi}} \sim g^2 \; \bar\psi_+\psi_+  \bar\psi_- \psi_-.
\eeq
This gauge invariant four-fermi coupling preserves the anomalous $U(1)_A$ symmetry, and is hence   expected  to be induced by perturbative loop effects in the bulk theory; we have indicated this by including the $g^2$  bulk coupling  in its definition. Thus, in the high-$T$ limit, we expect that the coefficient of this term is small, but have not calculated it precisely. The results of \cite{Sachs:1995dm}
for the gauged single-flavor Thirring model, with Lagrangian given by  (\ref{vectorDW}) and (\ref{4fermi}),  show   that it has the spectrum and chiral condensate of the massless Schwinger model, with the 4-fermi coupling inducing renormalization of  the boson mass and condensate. These effects are small  in the limit of small four-fermi coupling and we assume that, in our DW theory, they are negligible as $\beta \rightarrow 0$.

In addition to $U(1)_A$-preserving perturbative effects leading to (\ref{4fermi}), 
 bulk nonperturbative effects are known to  generate 't Hooft vertices. These  preserve only the $\mathbb Z_{4}^{d \chi}$ subgroup of $U(1)_A$. However, when projected on the DW worldvolume, they only induce higher-dimension non-renormalizable terms that are irrelevant from 2d perspective, in addition to being exponentially suppressed   in the high-$T$ phase. This follows already from the fact that (\ref{4fermi}) is the only gauge invariant and 2d Euclidean invariant local four-fermi coupling in the theory
   (\ref{vectorDW}).

Thus, borrowing the results of  Section \ref{Discrete 't Hooft anomalies}, we conclude that the DW theory breaks both the worldvolume center symmetry (originating from the $1$-form $\R^3$-bulk center symmetry) and the discrete chiral symmetry. This symmetry realization has some interesting implications:
\begin{enumerate}
\item  The broken $\mathbb Z_{4}^{d \chi}$ chiral symmetry implies that the fermion bilinear condensate should be nonzero on the DW in the high-$T$ (chirally restored!) phase, something that should be in principle measurable on the lattice. 

As far as the chiral symmetry is concerned, the high-$T$ dynamics on the DW  mirrors  the low-$T$ dynamics of the bulk, where  SYM  theory is known to have two vacua with a broken discrete chiral (or $R$-) $\mathbb Z_{4}^{d \chi}$ symmetry. 
\item The broken $\mathbb Z_2^C$ one-form center symmetry on the high-$T$ DW  implies a perimeter law for a fundamental Wilson loop taken to lie in the DW worldvolume. In contrast, recall that  a Wilson loop in the $\R^3$ bulk away from the DW exhibits area law. 

Here, the DW theory again reflects properties of the low-$T$ phase: the different behavior of the Wilson loop in the bulk and on the DW
 mirrors the deconfinement of quarks on the DWs (i.e. perimeter law for the Wilson loop along the DW) between chiral-breaking vacua in the confined low-$T$ phase (i.e. area law in the bulk), as found in \cite{Anber:2015kea}.
\end{enumerate}
The above relations between low-$T$ bulk physics and high-$T$ DW physics are quite tantalizing. One can not help but wonder about their generality and speculate on their  possible utility in constraining the features of   difficult to study strongly-coupled low-$T$ phases from the often more easily tractable  high-$T$ DW properties.  
\section{Outlook: generalizations and lattice studies }\label{proposed}

{\bf \flushleft{Higher} $\mathbf{n_f}$:} Motivated by the last remark in Section \ref{su2dw} we discuss the  higher-$n_f$ generalization of our study. Theories with different number  of adjoint flavors have been studied on the lattice with various motivations (a few recent references are \cite{Athenodorou:2014eua,Ali:2018dnd,Bergner:2018unx})  and a conclusive picture of their phase structure as a function of $n_f$ in the chiral limit has not emerged yet.  

As already alluded to, study of the $n_f >1$ high-$T$  DW theories requires more work and we only give a brief qualitative discussion. The DW fermion zero modes now come in $n_f$ multiples of those in (\ref{vectorDW}), which we label as  $\psi_{+, a}$, $\psi_{-, j}$, with  $a,b$ and $j,i$ denoting $SU(n_f)_{L,R}$ indices, respectively. The multi flavor generalization of the minimal coupling DW Lagrangian (\ref{vectorDW}) now has a $SU(n_f)_L \times SU(n_f)_R$ global symmetry, while the bulk QCD(adj) theory only has the $SU(n_f)$ global chiral symmetry. In the absence of  interactions other than those in (\ref{vectorDW}), the 't Hooft anomalies of the additional nonabelian global symmetries on the worldvolume have to be matched and one expects that new massless degrees of freedom on the DW worldvolume would appear.\footnote{In fact, the solution of the multiflavor massless Schwinger models in \cite{Kutasov:1994xq} shows that the massive spectrum of the multiflavor model is universal while a nonuniversal  massless sector matches the nonabelian flavor 't Hooft anomalies.} However, similar to (\ref{4fermi}), four-fermi terms are allowed and will be induced by bulk loop effects on the DW worldvolume, reducing    $SU(n_f)_L \times SU(n_f)_R$ to the diagonal subgroup, the bulk $SU(n_f)$ chiral symmetry. These perturbatively induced classically marginal terms should preserve the Euclidean rotations,  gauge symmetry, and anomalous chiral $U(1)_A$, and hence are expected to have the form
\beq
\label{4ferminf}
 {\cal L}_{DW}^{4-{\rm fermi}} \sim   \bar\psi_{+}^a \psi_{+, b} \; \bar\psi_-^i \psi_{-,j}\left(\alpha_1 \;\delta_a^b \;\delta_i^j +  \alpha_2\; \delta_a^j \;\delta^b_i\right)~.
 \eeq
 The $\alpha_2$ term  reduces the enhanced global symmetry of the free-fermion DW Lagrangian to the bulk $SU(n_f)$ chiral symmetry.\footnote{We note that for $n_f$$=$$2$ one can also write an invariant using $\epsilon^{bj} \epsilon_{ai}$ and stress that we have neither calculated nor studied the effect of any of the bulk-induced four-fermi terms. We also note that recently Ref.~\cite{Cordova:2018acb} pointed out subtleties related to additional  't Hooft anomalies for $n_f=2$.}  Thus, as opposed to the $n_f =1$ four-fermi term (\ref{4fermi}), we expect that  (\ref{4ferminf}) will affect the IR spectrum and possibly  the phase structure and symmetry realization (as far as terms breaking $U(1)_A$ but preserving $\mathbb Z_{4 n_f}$, the same comment as in the $n_f=1$ case applies---no such local terms relevant in 2d sense are allowed). 
 The higher-$n_f$ DW worldvolume theories have, as for $n_f = 1$, a mixed $\mathbb Z_{4 n_f}$-$\mathbb Z_{2}^C$ 't Hooft anomaly, exactly as in the bulk theory, while the $SU(n_f)$ global symmetry preserved by (\ref{4ferminf}) has no (mixed) anomaly in 2d. 
 
 We expect  that the multi flavor model with (\ref{4ferminf}) added will break the discrete chiral and center symmetries to match the anomaly. In light of the observed (so far, for $n_f =1$) tantalizing similarities between the behavior of the high-$T$ domain wall worldvolume theory  and the low-$T$ bulk theory, it would be worthwhile to study this further. It would be especially interesting to see whether there is any $n_f$ dependence of the DW worldvolume symmetry realization mirroring that of  the low-$T$ bulk theory.

An extension of our studies to higher numbers of colors would also be of interest.  

{\bf {\flushleft{Possible lattice}} studies:} One of the more surprising indications of our analysis  is that there should be a  nonzero fermion condensate and perimeter law for a (necessarily spacelike) Wilson loop in the  worldvolume of the high-$T$ DWs---all of this in the deconfined, chirally symmetric phase. 

 These effects should be, at least in principle,  observable in lattice studies. As lattice simulations are always performed at finite fermion mass, a chiral limit would have to be approached. While this is a difficult task, at least there is no sign problem (in the continuum limit, as opposed to $\theta=\pi$ pure gauge theories) for real values of the fermion mass. DW backgrounds in the high-$T$ phase can be induced by imposing appropriate twisted boundary conditions. For example, inserting a   $\mathbb Z_2^C$ phase in the action of a single gauge plaquette in the $x_1$-$x_4$ plane ($x_4$$\equiv$$x_4 + \beta$), for all   $x_2$ and $x_3$, would,  in the center-broken phase, induce  a DW  with an $x_2$-$x_3$ worldvolume (such configurations are also known as   high-$T$  center vortices \cite{Greensite:2011zz}; in this language, our studies amount to saying that they  have  a rather rich structure in theories with adjoint fermions).
 Needless to say, we leave the judgment regarding the feasibility of such studies to lattice experts. We only note that our analytic studies can be generalized to account for a small nonzero fermion mass. For example, to leading order in the fermion mass $m$, the degeneracy of the  $|P,\theta \rangle$ vacua (\ref{thetap}) in the charge-$q$ Schwinger model is lifted,  $E(P) \sim C' |m| \cos (\theta + {2 \pi P \over q})$, except for $\theta=\frac{\pi}{q}$, where $\theta$ now includes the phase of $m$.

{\bf {\flushleft{Other}}  theories with center symmetry:} Our final remark concerns the effects of mixed discrete chiral/center symmetry anomalies in other gauge theories. While theories with light or massless fundamental fermions have no  (not even approximate) center symmetry, there has been recent interest in theories with $n_f$ two-index symmetric (or antisymmetric) tensor Dirac fermions (see \cite{Fodor:2017rro} and references therein).  Other theories with center symmetry are discussed in \cite{Anber:2017pak}. The two-index symmetric (antisymmetric) tensor theories, with even number of colors,  have a $\mathbb Z_2^C$ $1$-form center symmetry and an anomaly free $\mathbb Z_{n_f(2 N+4)}^{d \chi}$ ($\mathbb Z_{n_f(2 N-4)}^{d \chi}$) $0$-form discrete chiral symmetry with a mixed 't Hooft anomaly. It would be interesting to study the matching of this anomaly in the various phases of these theories.

{\bf {\flushleft{Acknowledgments:}}} MA is supported by an NSF grant PHY-1720135 and the Murdock Charitable Trust. EP is supported by a Discovery Grant from NSERC.

  \bibliography{DWsu2biblio.bib}
  
  \bibliographystyle{JHEP}
  \end{document}